\documentclass[lettersize,journal]{IEEEtran}

\usepackage[utf8]{inputenc}

\usepackage[para,online,flushleft]{threeparttable}
\usepackage[english]{babel}
\usepackage{amsthm}
\usepackage{amsmath}
\usepackage{varwidth}
\usepackage{subcaption}
\usepackage{calc}
\usepackage{blindtext}
\usepackage{multicol}
\usepackage{graphicx}
\usepackage{paralist}
\usepackage[english]{babel}
\usepackage{float}
\usepackage{makecell}
\usepackage{multirow}
\usepackage{amsfonts}
\usepackage{booktabs}
\usepackage{siunitx}
\usepackage{etoolbox}
\usepackage{numprint}
\usepackage[T1]{fontenc}
\usepackage{algorithm}
\usepackage{flushend}
\usepackage[noend]{algpseudocode}
\usepackage{textcomp}
\usepackage{xcolor}
\usepackage{fancyvrb}
\usepackage{paralist}
\usepackage[inline]{enumitem}
\usepackage{booktabs}
\usepackage{color}
\usepackage{colortbl}
\usepackage{flexisym}
\usepackage{array}
\usepackage{xparse}
\usepackage{xspace}
\usepackage[flushleft]{threeparttable}
\usepackage[nobreak]{mdframed}
\usepackage{framed}
\usepackage{pifont}

\newcommand{\cmark}{\ding{51}}%
\newcommand{\xmark}{\ding{55}}%

\usepackage{listings}
\usepackage[]{xcolor}

\theoremstyle{definition}

\newcommand{\toolname}{\textsc{CigaR}\xspace}
\newcommand{\chatrepair}{\textsc{ChatRepair}\xspace}

\newcommand{\defectsfj}{\textsc{Defects4J}\xspace}
\newcommand{\humanevaljava}{\textsc{HumanEval-Java}\xspace}
\newcommand{\openai}{\textsc{OpenAI}\xspace}
\newcommand{\fixedbugscnt}{\textsc{171}\xspace}
\newcommand{\fixedbugsprcent}{\textsc{39.8\%}\xspace}
\newcommand{\totalbugscnt}{\textsc{429}\xspace}

\newcommand{\mycomment}[1]{}

\usepackage[]{xcolor}
\newcommand{\TODO}[1]{\textcolor{red}{#1}\GenericWarning{}{LaTeX Warning: TODO: #1}}\newcommand\todo\TODO

\newcommand{\MODIFIED}[1]{\textcolor{black}{#1}}\newcommand\modified\MODIFIED

\newcommand{\GHilight}{\makebox[0pt][l]{\color{green!20}\rule[-4pt]{1\linewidth}{10pt}}}

\newcommand{\RHilight}{\makebox[0pt][l]{\color{pink}\rule[-4pt]{1\linewidth}{10pt}}}
\newcommand{\YHilight}{\makebox[0pt][l]{\color{yellow}\rule[-4pt]{1\linewidth}{10pt}}}
\newcommand{\GRAYHilight}{\makebox[0pt][l]{\color{lightgray}\rule[-4pt]{1\linewidth}{10pt}}}

\algnewcommand{\Inputs}[1]{%
  \Statex \textbf{Inputs:}
  \Statex \hspace*{\algorithmicindent}\parbox[t]{.8\linewidth}{\raggedright #1}
}
\algnewcommand{\Outputs}[1]{%
  \Statex \textbf{Outputs:}
  \Statex \hspace*{\algorithmicindent}\parbox[t]{.8\linewidth}{\raggedright #1}
}
\algnewcommand{\Initialize}[1]{%
  \State \textbf{Initialize:}
  \Statex \hspace*{\algorithmicindent}\parbox[t]{.8\linewidth}{\raggedright #1}
}

\definecolor{javared}{rgb}{0.6,0,0} 
\definecolor{javagreen}{rgb}{0.25,0.5,0.35} 
\definecolor{javapurple}{rgb}{0.5,0,0.35} 
\definecolor{javadocblue}{rgb}{0.25,0.35,0.75} 

\lstdefinestyle{diff}{
    escapechar=\%
}

\usepackage{hyperref}
\addto\extrasenglish{%

}

\usepackage{cleveref}

\begin{document}

\title{CigaR: Cost-efficient Program Repair with LLMs}

\author{
Dávid Hidvégi$^*$,
Khashayar Etemadi$^*$,
Sofia Bobadilla,
Martin Monperrus%

\IEEEcompsocitemizethanks{
\IEEEcompsocthanksitem D. Hidvégi, K. Etemadi, S. Bobadilla, and M. Monperrus are with the KTH Royal Institute of Technology, Stockholm, Sweden\protect\\
Email: \{dhidvegi, khaes, sofbob, monperrus\}@kth.se\protect\\
}
}

\IEEEtitleabstractindextext{
\begin{abstract}
Large language models (LLM) have proven to be effective at automated program repair (APR). However, using LLMs can be costly, with companies invoicing users by the number of tokens. In this paper, we propose \toolname, the first LLM-based APR tool that focuses on minimizing the repair cost. \toolname works in two major steps: generating a first plausible patch and multiplying plausible patches. \toolname optimizes the prompts and the prompt setting to maximize the information given to LLMs using the smallest possible number of tokens. \modified{Our experiments on \totalbugscnt bugs from the widely used \defectsfj and \humanevaljava datasets shows that \toolname reduces the token cost by 73\%.
On average, \toolname spends 127k tokens per bug while the baseline uses 467k tokens per bug. On the subset of bugs that are fixed by both, \toolname spends 20k per bug while the baseline uses 608k tokens, a cost saving of 96\%.}
Our extensive experiments show that \toolname is a cost-effective LLM-based program repair tool that uses a low number of tokens to automatically  generate patches.
\end{abstract}

}

\maketitle

\IEEEdisplaynontitleabstractindextext

\IEEEpeerreviewmaketitle

\def\thefootnote{*}\footnotetext{These authors contributed equally to this work.}\def\thefootnote{\arabic{footnote}}

\section{Introduction}
Automated program repair (APR) \cite{monperrus2018automatic} aims to reduce the huge cost of software maintenance \cite{seacord2003modernizing} by automatically fixing bugs once they are detected. Automatizing software tasks is computationally costly, and it is no exception for APR research. \modified{Hence, it is important to make our APR systems as efficient as possible, and indeed it is a long standing problem in program repair to be able to fix bugs at a low cost \cite{LeGoues2012,xia2023keep}.} Recently, large language models (LLMs) have shown highly promising state-of-the-art results in APR \cite{jiang2023impact}. However, both training LLMs and using them for inference is costly \cite{chen2023frugalgpt}. For example, ChatGPT, which is arguably the most powerful existing LLM, charge their users for every single token that is either sent to the model or generated from it \cite{openaicost}.
In this paper, we study the problem of minimizing the computational cost of LLM-based program repair. 

\modified{In the context of LLMs, a cost-effective LLM-based APR tool should steer the model towards the correct patch with the lowest possible cost, i.e. with the lowest possible number of tokens. }For this purpose, two main components should be optimized: 1) the input to the LLM, i.e. the prompt, and 2) the configuration used for calling the model. First, a good prompt should maximize the trade-off between conciseness and information. While APR researchers have proposed prompt engineering techniques to better guide the model to find the correct patch \cite{zhang2023critical}, none of them have been optimized. Second, the settings under which the LLM works should lead to maximize the likelihood to output a correct patch, given a fixed token cost. Previous works have demonstrated that LLM settings, such as number of samples and temperature, are key to  obtain acceptable results with a low cost \cite{chen2023frugalgpt}. 

In this paper, we propose \toolname, a novel LLM-based program repair system that concentrates on token cost minimization. \toolname achieves cost-effectiveness with the help of its three delicately designed prompts working in concert: an `initiation prompt', an `improvement prompt', and a `multiplication prompt'. The initiation prompt initializes the repair process. The improvement prompt improves partial patches until a plausible patch is generated, avoiding throwing away potentially useful patches, hence avoiding wasting tokens. Finally, the multiplication prompt builds upon the already generated plausible patches to synthesize more plausible patches with diversity maximization. All these prompts are designed to be concise while staying informative, minimizing the overall token cost. The prompts help the LLM avoid unnecessary token cost by building upon its previous responses. \toolname also uses a reboot strategy to allow the model to look into various parts of its search space, instead of spending tokens in dead-ends.

\modified{We evaluate \toolname on two widely used bug datasets: \defectsfj \cite{just2014defects4j} and \humanevaljava. \toolname is highly effective, and possibly achieves the highest performance ever: \toolname is the first tool that fixes \fixedbugscnt/\totalbugscnt (\fixedbugsprcent) of the bugs in the considered dataset.} This shows that \toolname is properly taking advantage of its advanced prompt techniques for program repair. \modified{We demonstrate that \toolname reduces the token cost of the repair process by 73\%.} The significant cost reduction in \toolname proves the feasibility and effectiveness of the token cost minimization methods used by \toolname.
\modified{Given that energy consumption by LLMs is significant \cite{luccioni2023estimating}, saving 73\% of resources matters beyond program repair: it contributes to engineering frugal software tools in our era of climate crisis.}

To sum up, \toolname is the first LLM-based APR tool that aims at minimizing the computational cost, as measured by the number of tokens employed. \toolname is a sophisticated system that enables the LLM to find diverse plausible patches by:
1) summarizing the feedback given to the LLM as a part of an iterative approach;
2) rebooting the repair process after a few failing LLM invocations to allow the LLM look at various parts of its search space;
3) employing the LLM  to multiply the already generated patches in order to maximum diversity.

To summarize, we make the following contributions:
\begin{itemize}
    \item We introduce \toolname, a new LLM-based program repair approach. \toolname uses advanced prompting incl. iterative prompting, search rebooting and patch multiplication. \toolname has been designed and engineered to explore the patch search space effectively (more repaired bugs), while minimizing the repair token cost (lower invoice). 
    \item \modified{We evaluate \toolname on the widely used \defectsfj dataset, as well on \humanevaljava for sake of external validity. \toolname outperforms the state-of-the-art APR tools, incl. recent LLM-based APR tools, by fixing \fixedbugscnt/\totalbugscnt (\fixedbugsprcent)  of the considered bugs.}
    \item \modified{We study the amount of token cost reduction by \toolname. Compared to the state-of-the-art LLM-based APR, \toolname reduces the token cost by 73\% on average.} This shows the significance of token cost reduction by \toolname.
    \item We make \toolname publicly available for future research at \url{https://github.com/ASSERT-KTH/cigar}.    
\end{itemize}

The rest of this paper is organized as follows. In \autoref{sec:background}, we review the core concepts related to our research. In \autoref{sec:approach}, we explain how \toolname works. \autoref{sec:methodology} and \autoref{sec:results} present the methodology that we use for our experiments and their results. \autoref{sec:threats} discusses the threats to the validity of our results. In \autoref{sec:related_work}, we review the related work. Finally, in \autoref{sec:conclusion}, we conclude this paper.

\section{Background}
\label{sec:background}
In this section, we review the fundamental considerations and techniques that \toolname is built upon.

\subsection{Large Language Models}

Large language models (LLMs) are machine learning models with tens of billions of parameters trained on vast amounts of data. These models usually use the Transformer architecture \cite{vaswani2017attention} and consist of multiple fully connected layers of attention. LLMs can generate human-like text and code, they perform various natural language processing tasks such as translation, summarization, and programming code generation \cite{thirunavukarasu2023large,chen2021evaluating}.

There are three major types of Transformer models: encoder-only models, like BERT \cite{devlin2018bert}, decoder-only models, such as GPT-3 \cite{zhang2021commentary} \cite{umapathi2023med}, and finally encoder-decoder models like T5 \cite{raffel2020exploring}. Recent studies show that sizable decoder-only models outperform state-of-the-art in various software engineering tasks, including automatic program repair \cite{xia2023keep}.

A decoder-only LLM's procedure is as follows \cite{roberts2020much}. The user enters a textual prompt. Next, the prompt is tokenized and fed into the model. The model predicts the next token that should follow the tokenized input. This output token is then concatenated with the input tokens and sent again into the LLM in a loop until a stop condition is reached. The stop condition is a specific stop token or having exceeded the context length, which is the maximum number of tokens the model can be fed into. Tokens are the fundamental components based on which the model receives the input and generates the output. It is also the primary unit for counting usage and billing.

\subsection{LLM Token Cost}
LLMs take an input prompt made of a certain number of tokens, and then infer what tokens should follow the given input. All token in the input as well as the tokens generated by inference induce computational and financial costs for users \cite{cheng2023batch}.
\modified{For example, the GPT-3.5 Turbo model, which is one of the most advanced LLMs, charges \$0.0015 per 1k input tokens and \$0.002 per 1k generated token \cite{GPT_pricing} at the time of writing.} Overall, the computational cost of using these models is a growing concern in society due to environmental concerns and in research due to  the pressure on open, academic research \cite{wang2023cost}.

This motivates the study of the token-efficiency of LLM-based techniques.
The state-of-the-art metric for capturing token efficiency is the \emph{pass@t} metric  \cite{olausson2023demystifying}. This metric shows the likelihood that a model produces a correct answer with a token cost of \textit{t} tokens.
Techniques are employed to reduce the token costs, such as using less expensive models \cite{oppenlaender2023mapping},  hyperparameter optimization \cite{wang2023cost}, and in-context learning \cite{cheng2023batch}.

\section{\toolname Approach}
\label{sec:approach}

\subsection{Overview}
\label{sec:overview}

We design and implement \toolname, an LLM-based program repair tool that aims at minimizing the token cost. \toolname can be used with any LLM that has an API. \toolname employs three kinds of search:
1) search for a plausible patch from the buggy program, see \autoref{sec:initial_first_patch};
2) search for a plausible patch  from a partial patch, see \autoref{sec:subsequence};
3) search for more plausible patches from a single plausible patch, see \autoref{sec:alternative}.

\begin{figure*}
\begin{center}
\includegraphics[width=0.85\textwidth]{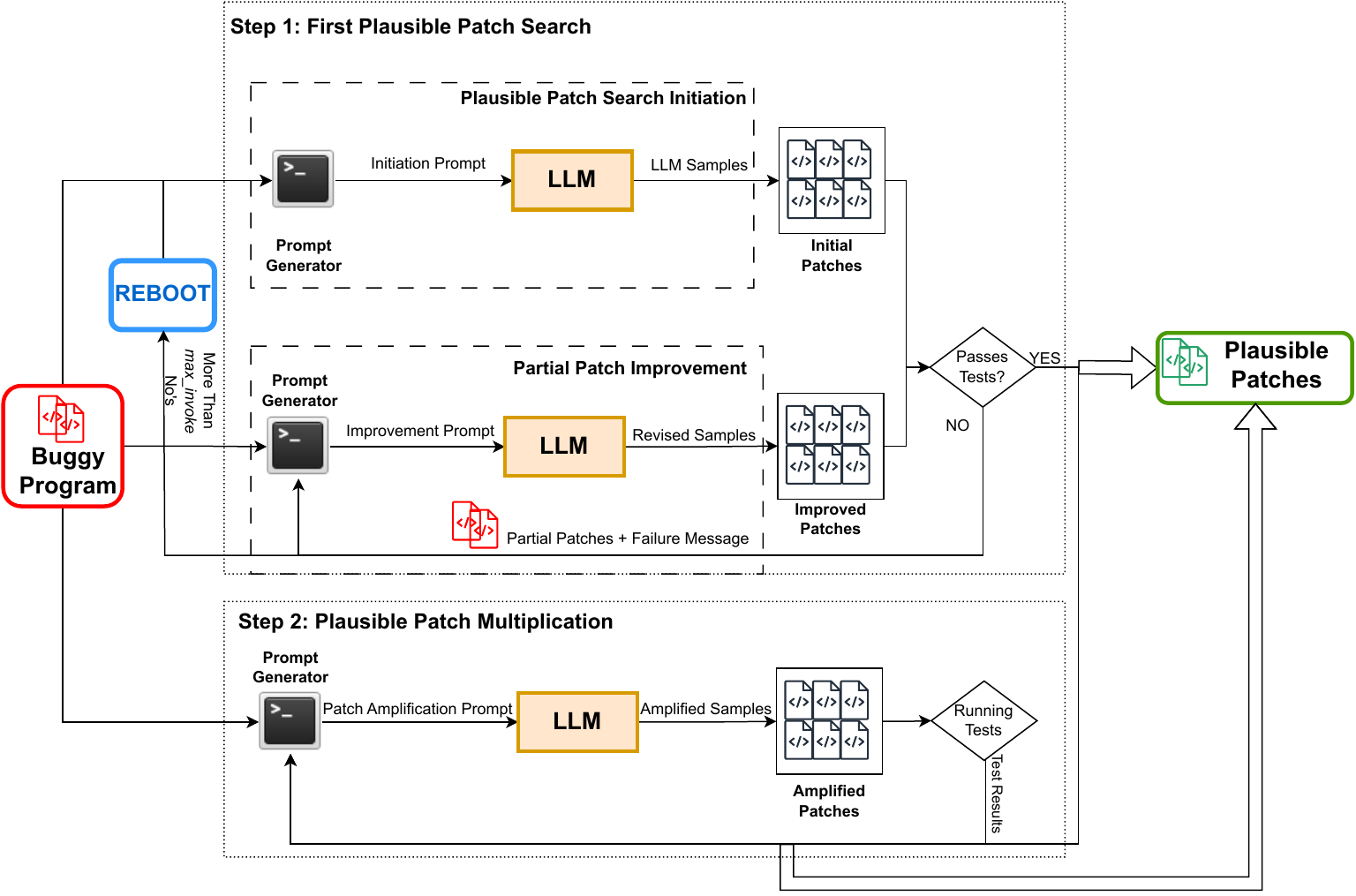}
\caption{Overview of \toolname: \toolname combines original prompting techniques, incl. partial patch improvement, search rebooting and patch multiplication to reach high effectiveness at a low token cost.}
\label{fig:rapidcapr}
\end{center}
\end{figure*}

\autoref{fig:rapidcapr} represents an overview of how \toolname works.
\toolname takes as input a buggy program that fails on a test  and generates a set of plausible patches that pass all tests as the output. \toolname works in two major steps, ``First Plausible Patch Search'' and ``Plausible Patch Multiplication''. In the first plausible patch search step, \toolname tries to generate a plausible patch that passes all the tests. For this, \toolname uses so-called `initiation prompt'. If the LLM fails to suggest a plausible patch in its first invocation, \toolname proceeds by using improvement prompts. In the plausible patch multiplication step, after the first plausible patch is generated, \toolname asks the LLM to produce alternative plausible patches.

The token minimization approach of \toolname is based on three novel concepts:
1) the concept of patch multiplication, steering the LLM to generate maximally diverse plausible patches,
2) the design of three types of informative prompts which provide the LLM with a summarization of its responses in previous iterations,
3) the reboot of the repair process.

\subsection{Step 1: First Plausible Patch Search}
\subsubsection{Plausible Patch Search Initiation}
\label{sec:initial_first_patch}

\toolname starts with an initiation prompt, whose goal is to identify a first plausible patch. To get the information for the initiation prompt, \toolname first runs the tests on the buggy program and saves the test results. Next, a prompt is generated that consists of three core sections.
\begin{enumerate}
    \item A one-shot example of fixing a bug, to ensure that \toolname takes advantage of the in-context learning to understand the expected format.
    \item The buggy code, which is the main part of the input, per a fault localization algorithm.
    \item The test failure details, with valuable extra information about the bug under repair.
\end{enumerate}

\begin{figure}
\begin{center}
\includegraphics[width=0.5\textwidth]{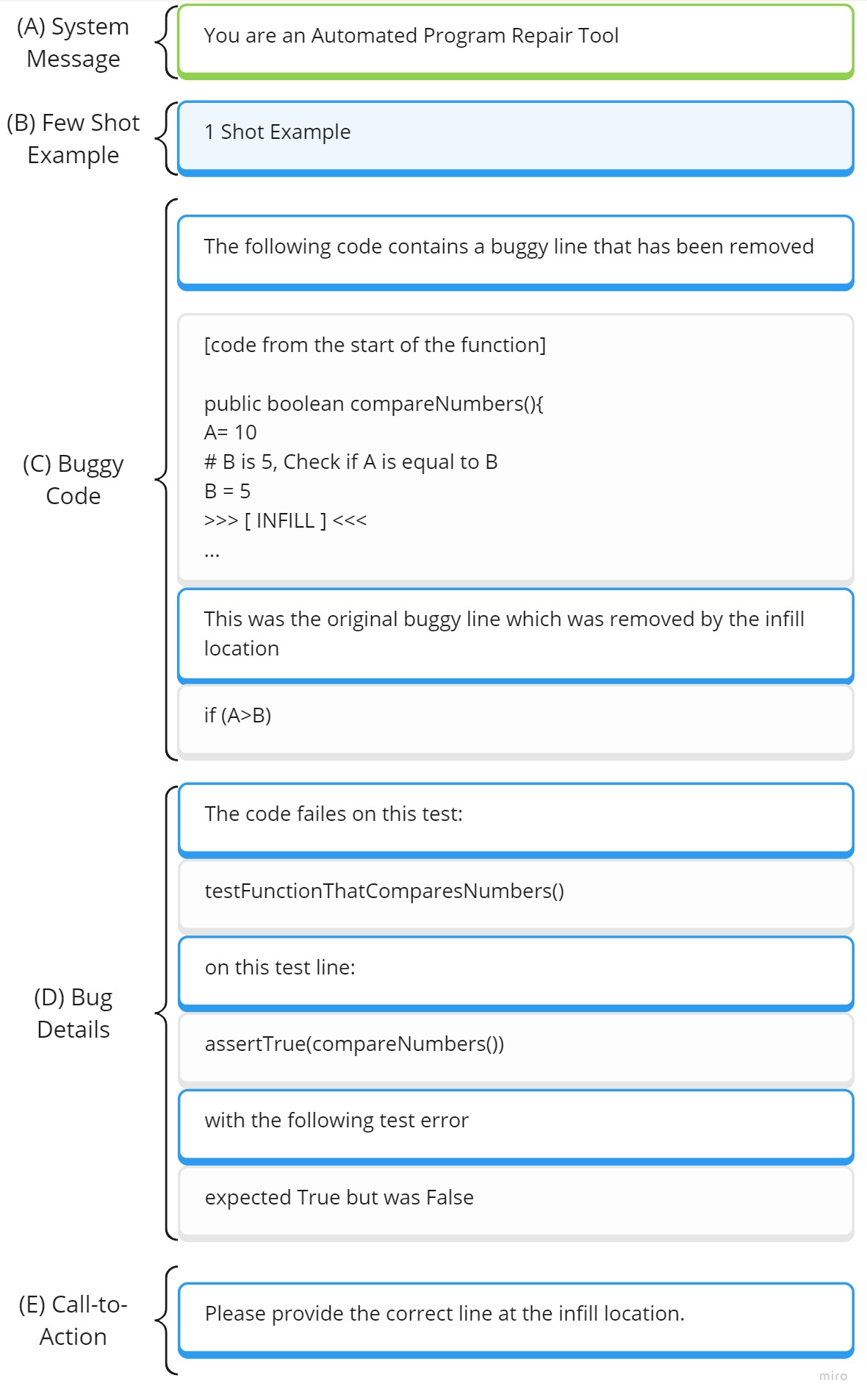}
\caption{P1: initiation prompt structure used for plausible patch search initiation.}
\label{fig:initial_prompt}
\end{center}
\end{figure}

\autoref{fig:initial_prompt} illustrates the structure of the initiation prompt employed by \toolname. At the beginning, there is a system message (A) to put the LLM in the role of an APR tool. The second part of the prompt is a one-shot example (B), which includes a small bug and its fix in the history of the same software project. We select the smallest bug for the one-shot example in line with our token minimization strategy.

The third part of the input is the buggy code (C). The buggy part of the code is replaced with an ``\texttt{[INFILL]}'' token, and the original buggy part of the code is separately presented to the model. Note that \toolname fixes bugs that are localized inside a single function.

The fourth part of the prompt presents the test failure details (D). This includes the failing test method, the assertion that triggers the failure, and the runtime error message. This gives the model extra information regarding the execution of the buggy code that leads to the failure.

Finally, there is a call to action section (E) in the prompt that specifies what we exactly expect the LLM to generate.

The initiation prompt is sent to the model and the LLM is asked to generate 50 samples. This is because we want to generate as many samples as possible with a single input token upfront cost. Per our preliminary experiments, generating 50 samples per request is the sweet spot that gives us the most diverse set of patches without generating too many repetitive responses which cost output tokens for nothing.

\toolname extracts the generated code from the response and substitutes the original buggy code for it. The result is a set of `initial patches'. These initial patches are tested. A patch that passes all the tests is a plausible patch, and a patch that does not pass all the tests is a partial patch. If there is any plausible patch among the initial patches, it is directly sent to the plausible patch multiplication step, explained in \autoref{sec:alternative}. Otherwise, the best partial patch together with its failure error message is sent to the partial patch improvement phase, explained in \autoref{sec:subsequence}. The best partial patch is the first patch without compilation errors.
If all generated patches have compilation errors, the first generated patch is considered as the best partial patch.

\subsubsection{Partial Patch Improvement}
\label{sec:subsequence}

The goal of partial patch improvement is to get a plausible patch by improving partial patches found during the plausible patch search initiation phase.
In this phase, \toolname takes an iterative approach for generating a plausible patch: until the model generates a plausible patch, \toolname invokes the LLM with improvement prompts. The improvement prompt contains a system message, the buggy code, test failure details, and a call to action similar to the initiation prompt as explained in \autoref{sec:initial_first_patch}.

In addition to the sections common with the initiation prompt, the improvement prompt also gives the LLM a concise feedback on its previous responses. This ensures that the model looks for the plausible patch at different parts of the search space with a low token cost. The information regarding the past generated partial patches comes right before the call to action. The partial patches generated in previous iterations are grouped by their failure message. Patches in each of these groups are added next to each other, and their common failure message comes after them. Using this method, we summarize previously generated patches without repeating test failure messages.
\toolname considers as many previous patches as it is possible in a single prompt without going beyond the LLM's prompt token limit. 

Note that compared to the initiation prompt, the one-shot example is removed in the subsequent prompt. The reason is that the one-shot example is mainly used to help the LLM grasp the style of output it should generate. In the improvement prompt, the previous implausible patches play the same role and hint the model in what format it should generate the output.

The LLM is reinvoked with improvement prompts as long as no plausible patch is generated, and we have not exceeded $max\_invoke$ invocations. $max\_invoke$ is a configurable parameter in \toolname, set to ten by default per our pilot experiment. As soon as an improvement prompt generates a plausible patch, \toolname sends it to the third phase (\autoref{sec:alternative}). Otherwise, \toolname has failed to fix the given buggy code.

\subsubsection{Rebooting the Plausible Patch Search}
In case the LLM fails to synthesize a plausible patch after $max\_invoke$ invocations, \toolname reboots the whole repair process. This means ignoring all the patches generated before and starting from the initiation prompt. Rebooting the repair process enables \toolname to explore an even larger repair space by starting from a different initial seed, according to the model temperature. For this, \toolname uses a high temperature (see \autoref{sec:impelementation}), which maximize exploration. The rebooting strategy with high temperature leads to initiating the repair process from a radically different hidden state, meaning exploring an undiscovered part of the repair space. As a part of our token minimization approach, this helps \toolname avoid using too many tokens for exploring a limited and fruitless part of the search space.

Every time \toolname reboots the repair process, we say it has started a new \emph{round}. Round\_1 is when \toolname starts the repair and round\_i is the repair process after the $(i-1)$th reboot.

\subsection{Step 2: Plausible Patch Multiplication}
\label{sec:alternative}

\begin{figure}[t]
\begin{center}
\includegraphics[width=0.5\textwidth]{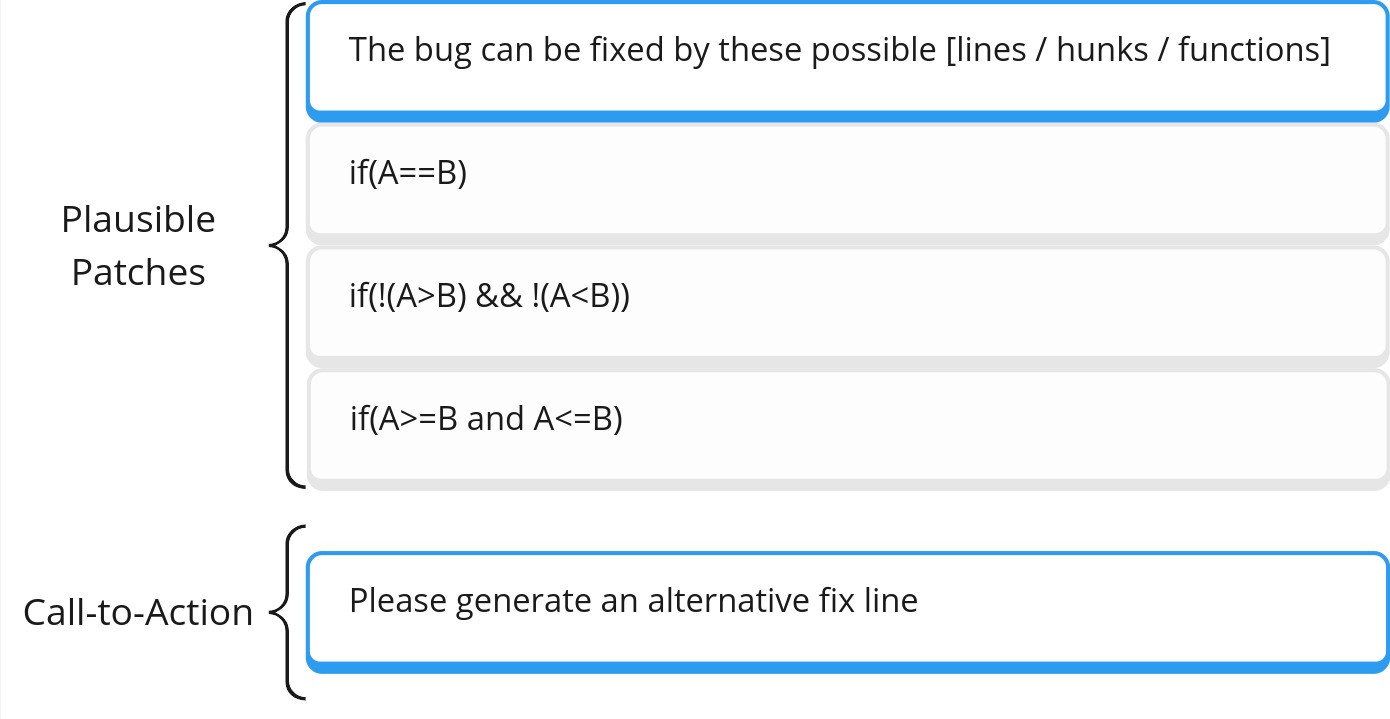}
\caption{P2: Prompt used for plausible patch multiplication.}
\label{fig:alternative_prompt}
\end{center}
\end{figure}

Plausible patch multiplication is the final step, its goal is to generate more distinct plausible patches.
After the first plausible patch is generated with an initiation or an improvement prompt, \toolname tries to generate new plausible patches. The reason is that the program repair literature has shown that even a plausible patch can be semantically incorrect \cite{ye2021automated}. By generating multiple plausible patches, \toolname has a better chance of producing a correct patch. \toolname can employ any patch ranking technique \cite{ghanbari2022patch} to select the most likely patch out of the set of plausible patches generated.

The patch multiplication process also works iteratively. In this step, the prompt contains the following parts: a system message and the buggy code similar to the initial prompt, as well as a summary of the generated plausible patches and a call to action to generate different patches. \autoref{fig:alternative_prompt} shows an example of these latter two parts. \toolname includes as many of the recent plausible patches as possible without exceeding the LLM's prompt token limit. This makes the LLM amplify the set of existing patches with patches that are distinct from the previous ones, with a reduced token budget. \toolname invokes the LLM with multiplication prompts and collecting the responses five times. We aim for five multiplication tries to explore a significant part of the plausible patch search space without generating many repetitive patches.

Finally, the full list of all generated plausible patches is reported to the user.

\subsection{Implementation}
\label{sec:impelementation}

\toolname is implemented in Python and  uses OpenAI ``\texttt{gpt-3.5-turbo\\-0301}'' as the underlying LLM with sampling temperature of 1. This high temperature adds a notable level of randomness to LLM's output. This is essential for exploring different parts of the repair space with \toolname's reboots and patch multiplication techniques. We also note that \toolname implements a thorough caching system, which stores all the prompts sent to the LLM, all responses received from the LLM as well as the test execution results. This caching system is important for reproducing and investigating past repairs, as invoking the LLM can be highly costly. \toolname and all the data related to our experiments are made publicly available\footnote{\url{https://github.com/ASSERT-KTH/cigar}}.

\section{Experimental Methodology}
\label{sec:methodology}

\subsection{Research Questions}

\newcommand\rqone{How effective is \toolname in terms of generating plausible and correct patches?}

\newcommand\rqtwo{How efficient is \toolname in terms of token cost compared to the state-of-the-art?}

\newcommand\rqthree{To what extent does patch exploration advance over reboots and multiplications?}


In this paper, we study the following research questions.
\begin{itemize}
    \item RQ1 (effectiveness): \rqone \ We count the number of bugs for which \toolname generates a plausible and correct patch, respectively. We compare \toolname against the state-of-the-art APR tools based on \openai's LLMs.
    \item RQ2 (efficiency): \rqtwo \ \modified{We run \toolname on \totalbugscnt real-world bugs from the widely used datasets, \defectsfj and \humanevaljava.} We compare \toolname's token efficiency against \chatrepair \cite{xia2023keep}.
    \item RQ3 (patch exploration): \rqthree \ We count the number of distinct patches that are the result of \toolname's iterative approach. As repeating the LLM invocations is the major source of \toolname's token cost, this studies whether our reboot and multiplication strategies are worthwhile.
\end{itemize}

\subsection{Study Subjects}
To evaluate \toolname, we need a benchmark. \modified{Therefore, we use \defectsfj and \humanevaljava \cite{jiang2023impact} as the benchmarks of our study. \defectsfj is the most commonly used benchmark for APR. It enables us to compare against performance metrics available in the published papers or their replication packages. We also consider \humanevaljava \cite{jiang2023impact} to address the potential data leakage concern. \humanevaljava is a benchmark designed fore being more recent than the training data of the \texttt{gpt-3.5-turbo-0301} model. Therefore, a good performance on \humanevaljava indicates our results are not susceptible to data leakage.}

\begin{table}[t]
\centering
\caption{\modified{Statistics for the considered bugs from the \defectsfj dataset.}}
\label{tab:defects4j_6_projects}
\begin{tabular}{@{} l ccccc @{}}
\toprule 
{Identifier} & {Project Name} & {All Bugs} & {Single-Function}  \\ [0.5ex] 
\midrule
Chart & jfreechart & 26 & 17 \\
Closure & closure-compiler & 174 & 102 \\
Lang & commons-lang & 64 & 42 \\
Math & commons-math & 106 & 73 \\
Mockito & mockito & 38 & 17 \\
Time & joda-time & 26 & 16 \\
{\small \humanevaljava} & -- & 162 & 162 \\
\midrule
\textbf{Total} & -- & \textbf{596} & \textbf{429} \\ 
\bottomrule
\end{tabular}
\end{table}

\modified{As \toolname fixes single-function bugs, we consider only bugs of this type. \autoref{tab:defects4j_6_projects} shows the characteristics of the projects in our dataset. The ``All Bugs'' column presents the total number of bugs from that project in the most recent version of \defectsfj and \humanevaljava.} The ``Single-Function'' column shows the number of single-function bugs.
\modified{Overall, our experiments consider 429 single-function bugs.} Note that each project contains at least 16 single-function bugs, which makes a diverse and appropriate dataset for our study.

\subsection{Protocol for RQ1 (Effectiveness)}
\label{sec:rq1_protocol}

\modified{To answer \textbf{RQ1}, we run \toolname on all \totalbugscnt bugs in our dataset.} After each round of executing \toolname ends, we identify the bugs for which a plausible patch is generated, and the bugs with no plausible patch. We continue rebooting \toolname until one plausible patch is found or the maximum number of round is reached.

We assess the effectiveness of \toolname by computing the number and ratio of bugs for which it generates a plausible or correct patch. 
We consider a patch to be plausible when it passes all the provided tests. 
We consider a patch to be correct if its abstract syntax tree (AST) exactly matches the AST of the ground-truth fix for the bug in our dataset. This allows for being oblivious to formatting changes. When a tool generates a correct patch for a bug, we say the tool fixes the bug. The more correct/plausible patches generated by \toolname, the more effective it is. 
Note that generating a maximum number of distinct plausible patches is good, because it maximizes the underlying of having the correct one. Techniques such as patch ranking \cite{ye2021automated} are then responsible for finding the correct one.

We also compare \toolname against LLM-based APR tools that employ \openai's GPT models and were evaluated on  \defectsfj bugs. To obtain a full list of such tools, we search for ``GPT'' and related LLM names in the Google Scholar academic search engine. We then manually check all the papers and identify the ones that run their proposed tool on all applicable \defectsfj bugs and report the number of correct patches.
The resulting set includes \emph{nl2fix}~\cite{fakhoury2023towards}, \textsc{STEAM}~\cite{zhang2023steam} and \chatrepair. 
\emph{nl2fix} augments the prompt with issue descriptions.
\textsc{STEAM} takes an interactive approach with multiple steps to fix \defectsfj bugs. 
\chatrepair \cite{xia2023keep} has an iterative approach and gives test results to the LLM as feedback. \modified{For the sake of comparison, we extract the numbers reported in the original paper of \emph{nl2fix} and \textsc{STEAM}}.

Per our literature review, \chatrepair is the most  advanced APR tool that uses OpenAI's models. Therefore, having the actual patches generated by \chatrepair and its prompts is essential for the study of \toolname's effectiveness and efficiency against the state-of-the-art. As \chatrepair and its patches for \defectsfj are not publicly available, by October 4, 2023, we carefully reimplement it according to the paper, incl. private communication with the authors. Our reimplementation of \chatrepair is publicly available in our repository for researchers and practitioners.

In our comparison between \toolname and the state-of-the-art APR tools, we measure performance in terms of the ratio of bugs with a correct and plausible generated patch. 
Specifically for \toolname, we also track how the number of correct/plausible patches increases as \toolname enters new rounds. A significant increase shows that the effectiveness of the rebooting strategy adopted by \toolname.

We also check the overlap between the bugs that each of \toolname and \chatrepair fix. This experiment shows us whether some tools explore a different part of the repair search space. We only consider \toolname and \chatrepair for this study,  because the patches for \emph{nl2fix} and \textsc{STEAM} are not publicly available.

\subsection{Protocol for RQ2 (Efficiency)}
\label{sec:rq2_protocol}

In answer to RQ2, we compare \toolname against \chatrepair only, because it is the only tool for which we can precisely measure token usage. 
We compare \toolname's total token cost against the total token cost of \chatrepair. Moreover, we study the token cost of \toolname and \chatrepair on bugs that are fixed vs not-fixed by each tools per our RQ1 experimental results. This sheds light on a different aspect of the efficiency of each tool, as it splits the token cost to fruitful ones leading to a correct patch and unfruitful ones, where the token price to pay is a pure loss. We want to see if these two tools spend a significant number of unnecessary tokens on bugs that are not fixed at the end.

\subsection{Protocol for RQ3 (Exploration)}
\label{sec:rq3_protocol}

To answer RQ3, we measure how the number of distinct patches generated by \toolname increases, as it makes more LLM invocations. A patch \emph{p} for a buggy program \emph{b} is distinct if \emph{p} is textually different from all patches generated for \emph{b} in previous LLM invocations. Note that a distinct patch may be correct, plausible, partially plausible, or even uncompilable. Synthesizing a large number of distinct patches means \toolname is exploring different parts of the search space without repeating itself. Such patch synthesis is essential for making the exploration cost-effective and finding a correct patch with a low cost.

We count the total number of distinct generated patches as well as the number of distinct plausible patches for ``plausible patch multiplication'', which is in line with the actual target of this step. A consistent increase in the number of distinct generated (plausible) patches shows the patch exploration capabilities in \toolname's repair process.

We conduct this experiment on a set of selected bugs in our dataset. \modified{For studying the first plausible patch search step, we select the bugs that take the maximum number of LLM invocations to be plausibly repaired per each \defectsfj project in the dataset, which means they are very hard to fix, and much exploration is needed.}

For the plausible patch multiplication step, we make the set of selected bugs more diverse by also considering the bugs for which the first plausible patch is generated with the first LLM invocation. Note that at the plausible patch multiplication step, the number of LLM invocations is always set to five, which provides the same amount of data for all bugs.

By studying the number of distinct generated (plausible) patches relative to the number of LLM invocations, we better understand how \toolname explores the patches in the repair space for different types of bugs.
In particular, we study whether \toolname cause unnecessary token cost by repeating patches that have been synthesized in previous LLM invocations.

\section{Experimental Results}
\label{sec:results}

\subsection{Results for RQ1  (Effectiveness)}
\label{sec:results_rq1}

\begin{table*}[t]
\centering
\scriptsize
\caption{\modified{Effectiveness of \toolname vs recent LLM-based program repair tools on \defectsfj.}}
\label{tab:effectiveness}
\begin{tabular}{@{}l | r r r | r r r | r r r @{}}
    \toprule
    & \multicolumn{3}{c}{\defectsfj} & \multicolumn{3}{c}{\humanevaljava} & \multicolumn{3}{c}{ALL} \\
    Tool & \#Bugs & \#Plausible & \#Correct (EM) & \#Bugs & \#Plausible & \#Correct (EM) & \#Bugs & \#Plausible & \#Correct (EM) \\
    \midrule
    \emph{nl2fix} & 283 & 53.3\% (151/283) & 11.3\% (32/283) & -- & -- & -- & -- & -- & -- \\
    \textsc{STEAM} & 260 & -- & 25.0\% (65/260) & -- & -- & --  & -- & -- & -- \\
    \chatrepair & 267 & 59.9\% (160/267) & 19.8\% (53/267) & 162 & 93.2\% (151/162) & 52.4\% (85/162) & 429 & 72.4\% (311/429) & 32.1\% (138/429) \\
    \midrule
    \toolname\_R1 & 267 & 56.5\% (151/267) & 23.9\% (64/267) & 162 & 83.9\% (136/162) & 61.1\% (99/162) & 429 & 66.8\% (287/429) & 33.7\% (145/429) \\
    \toolname\_R3 & 267 & 61.4\% (164/267) & 25.4\% (68/267) & 162 & 90.1\% (146/162) & 62.9\% (102/162) & 429 & 72.2\% (310/429) & 39.6\% (170/429) \\
    \toolname\_R12 & 267 & 69.2\% (185/267) & 25.8\% (69/267) & 162 & 93.8\% (152/162) & 62.9\% (102/162) & 429 & 78.5\% (337/429) & 39.8\% (171/429) \\
	\bottomrule
\end{tabular}
\end{table*}

\modified{\autoref{tab:effectiveness} shows the results of running \toolname on our dataset and compares it with the state-of-the-art tools. The table presents the results for \defectsfj and \humanevaljava bugs both separately and combined.} The ``Tool'' column shows the name of the APR tool. As explained in \autoref{sec:rq1_protocol}, we reboot \toolname several times until it generates a plausible patch for one of the remaining unfixed bugs. This strategy leads to running \toolname for 12 rounds. The results for three different rounds is presented in \autoref{tab:effectiveness}. \modified{``\#Bugs'' represents the total number of bugs on which the APR tool is run.}

\modified{Note that the ``\#Bugs'' column that shows the number of considered \defectsfj bugs slightly varies between the tools.} This is due to the differences between the expectations of each tool. However, the overall number of considered bugs is similar between the considered tools, which enables us to have a meaningful comparison. 

\modified{The ``\#Plausible'' and ``\#Correct (EM)''  columns show the number of bugs for which the APR has generated a plausible and a correct patch, respectively. For example, row \toolname\_R3 shows that after running \toolname for three rounds, it generates a plausible patch for 61.4\% of \defectsfj bugs and 90.1\% of \humanevaljava bugs. In total, it generates a plausible patch for 72.2\% or 310 of \totalbugscnt considered bugs in our experiment. For 39.6\% (170/\totalbugscnt) of the bugs, one of the generated plausible patches is correct as it has an AST exactly matching the AST of the ground-truth fix. Note that the performance of \emph{nl2fix} and \textsc{STEAM} are not known for \humanevaljava and their tool is not publicly available. }

\modified{\toolname is able to generate a plausible patch for 69.2\% of \defectsfj bugs and 93.8\% of \humanevaljava bugs, which is arguably high. In total, \toolname produces plausible patches for 78.5\% (337/\totalbugscnt) of the bugs in our dataset. This is higher than the best performance by state-of-the-art, \chatrepair, which generates plausible patches for 59.9\% (160/267) of \defectsfj bugs and 93.2\% (151/162) of \humanevaljava. In the first round R1, \toolname generates a plausible patch for 66.8\% (287/\totalbugscnt) of the bugs, and this number increases at each round until the twelfth round.} This shows the effectiveness of the reboot strategy adopted by \toolname, which maximizes exploration  using a new random seed thanks to the temperature mechanism.

Generating a correct patch is the ultimate goal of any APR tool. \toolname outperforms all other tools by generating a correct patch for 25.8\% (69/267) of \defectsfj bugs and 93.8\% (152/162) of \humanevaljava bugs. This proves the overall end-to-end effectiveness of \toolname. \modified{The number of bugs with a correct patch also increases over rounds, again demonstrating the importance of rebooting. \toolname surpasses all existing tools on both \defectsfj and \humanevaljava in terms of correct patches at round three (R3).}

\begin{figure}
\begin{center}
\includegraphics[width=0.5\textwidth]{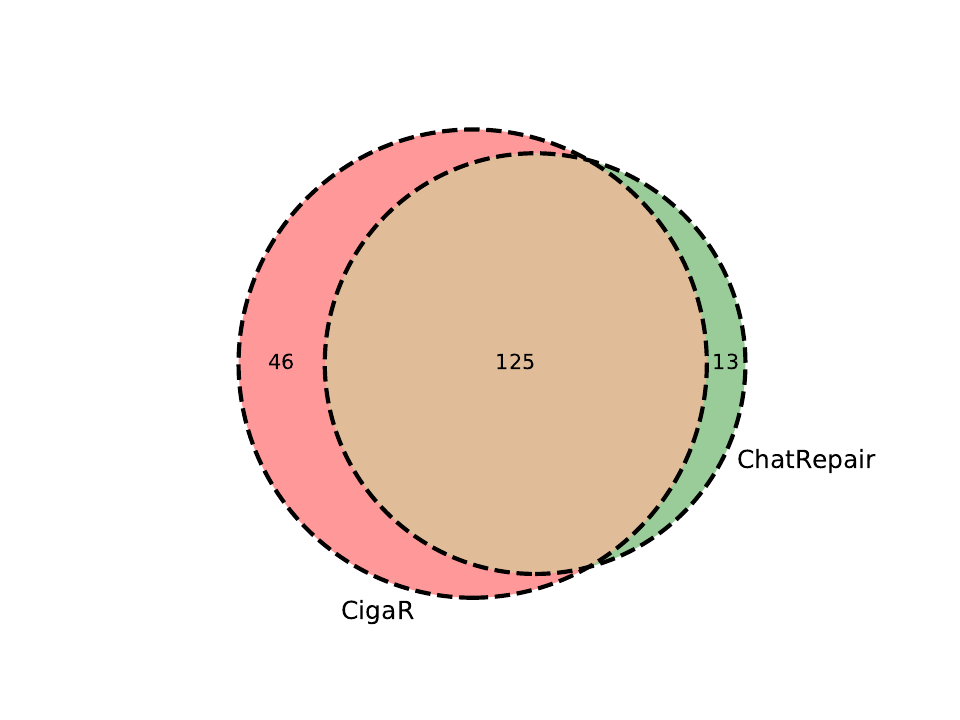}
\caption{\modified{The overlap between bugs fixed by \toolname and \chatrepair in our merge dataset (Defects4J and HumanEval).}}
\label{fig:rq1_overlap}
\end{center}
\end{figure}

\autoref{fig:rq1_overlap} illustrates the overlap between bugs that  \chatrepair and \toolname fix (we cannot do it for the other tools due to the absence of detailed information in the respective papers). \modified{There are 125 bugs that both tools generate a correct patch for. Moreover, 26\% (46/171) of the bugs fixed by \toolname are not fixed by \chatrepair, which shows \toolname's patch exploration strategy looks into parts of LLM's repair space that was not covered previously.}

\begin{mdframed}\noindent
    \textbf{Answer to RQ1: \rqone} \\
    Our experiment on \defectsfj shows that \toolname is effective at automatically fixing real-world bugs. \toolname generates plausible patches for 78.5\% (337/\totalbugscnt) bugs. It generates 39.8\% (171/\totalbugscnt)  correct patches with the strictest definition of correctness -- exact match with the developer ground truth patch. This outperforms the related APR tools built on \openai's GPT models, thus showing the potency of \toolname's prompting techniques.
    Notably, 46 bugs in the dataset are only fixed by \toolname. 
    The validity of those results is fully confirmed by our run on a new benchmark \humanevaljava, which is strong evidence of external validity.
    Overall, these experiments demonstrate that \toolname's novel iterative prompting pipeline effectively explore the search space of LLMs to find correct patches.
\end{mdframed}

\subsection{Results for RQ2 (Efficiency)}
\label{sec:result_rq2}

\begin{table}[t]
\centering
\footnotesize
\caption{\modified{Token cost of \toolname and \chatrepair on various types of bugs. Lower cost indicates better token cost efficiency.}}
\label{tab:token_cost}
\begin{tabular}{@{} l rrllr @{}}
\toprule 
\textbf{Bugs\_Fixed\_By} & \textbf{CHC} & \textbf{CIC} & \textbf{CHC_AVG} & \textbf{CIC_AVG} & \textbf{Saving} \\
\midrule
\chatrepair (13) & 8.6M & 1.5M & \colorbox{black!30}{\parbox{1.3em}{\hfill}} 661K & \colorbox{black!30}{\parbox{0.2em}{\hfill}} 115K & -- \\
\toolname (46) & 24.4M & 1.5M & \colorbox{black!30}{\parbox{1em}{\hfill}} 530K & \colorbox{black!30}{\parbox{0.05em}{\hfill}} \xspace \xspace 32K & -- \\
Both (125) & 76M & 2.6M & \colorbox{black!30}{\parbox{1.15em}{\hfill}} 608K & \colorbox{black!30}{\parbox{0.03em}{\hfill}} \xspace \xspace 20K & 96\% \\
Neither (245) & 95.3M & 49.1M & \colorbox{black!30}{\parbox{0.6em}{\hfill}} 388K & \colorbox{black!30}{\parbox{0.4em}{\hfill}} 200K & 48\% \\
\midrule
\textbf{Total} (429) & 204.3M & 54.9M & \colorbox{black!30}{\parbox{0.9em}{\hfill}} 467K & \colorbox{black!30}{\parbox{0.27em}{\hfill}} 127K & 73\% \\
\bottomrule
\end{tabular}
\end{table}

Next, we focus on the token cost. We only compare with \chatrepair as the implementations of \textsc{STEAM} and \textit{nl2fix} are not publicly available, preventing us for collecting their respective token cost.

We compute each tools' cost on bugs of different types. \autoref{tab:token_cost} shows the result of this experiment. The ``Bugs\_Fixed\_By'' column indicates the type of bugs according to the tool(s) that fix them. The number in brackets in this column is the number of bugs of the respective type. ``CHC'' and ``CIC'' show the token cost of \chatrepair and \toolname on each type of bug, respectively. Also, ``CHC\_AVG'' and ``CIC\_AVG'' represent \chatrepair's and \toolname's average token cost per bug for each type. Finally, the ``Saving'' column indicates the percentage of saved token cost by \toolname compared to \chatrepair. 

\modified{In total, \chatrepair spends 467K tokens on average, while \toolname spends 127K. This means \toolname improves the token cost by 73\% (149.4M/204.3M), while, per RQ1, it outperforms \chatrepair in terms of the number of plausible/correct generated patches, demonstrating a win-win situation.} This is explained by of our token minimization strategies:  asking for as many samples as possible at each LLM invocation, timely reboots, and summarizing the previous patches in the prompt. To sum up, \toolname improves on both aspects: 1) it enables us to have more distinct patches and 2) for fewer tokens.

\modified{In \autoref{tab:token_cost}, we see that \toolname saves 96\% (73.4M/76M) of token cost on bugs that are fixed by both. We also see that \toolname spends 32K tokens on average for the 46 bugs that are only fixed by \toolname, while \chatrepair spends 661K tokens on average for the 13 bugs only fixed by \chatrepair.} All this data shows, when the tool is able to find a correct patch, \toolname works much more efficiently than \chatrepair. By manually checking our experimental data, we see the reason is that \chatrepair uses up to 199 LLM invocations after generating the first plausible patch to produce alternative plausible patches one by one. However, \toolname only uses five LLM invocations for plausible patch multiplication and generates up to 50 patches at each invocation. We see that \toolname's summarization of previous patches in each invocation and using a high temperature is effective: it leads the LLM to synthesize novel patches without using too many invocations and causing unnecessary cost.

\begin{lstlisting}[float=tb, style=diff, caption={The correct patch for the bug ``Mockito\_12'', which is fixed by both \chatrepair and \toolname. \chatrepair spends 275K tokens on this bug, while \toolname spends 22K saving 92\% of the token cost.}, captionpos=b, label=lst:cost_ex,belowskip=-0.4\baselineskip]
  Type actual = ((ParameterizedType) generic).getActualTypeArguments()[0];
%\GHilight%+ if (actual instanceof Class) {
  return (Class) actual;
%\GHilight%+ } else if (actual instanceof ParameterizedType) {
%\GHilight%+   return (Class) ((ParameterizedType) actual).getRawType();
%\GHilight%+ }
\end{lstlisting}

\autoref{lst:cost_ex} shows \modified{the bug ``Mockito\_12''} and its fix by both \toolname and \chatrepair. In the buggy version, the program always casts the variable \texttt{actual} to a \texttt{Class} and returns it. In the fixed version generated by \toolname and \chatrepair, it is first checked whether \texttt{actual} is an instance of \texttt{Class}; if not, the raw type of \texttt{actual} is returned. On this bug, \toolname saves 92\% (253K/275K) of the token cost, compared with \chatrepair . Our manual analysis on this bug confirms our claim regarding cost saving on bugs fixed by both tools: \chatrepair generates alternative plausible patches one-by-one with up to 199 LLM invocations, while \toolname uses only 5 LLM invocations with a sample size of 50.

\modified{Finally, we consider the 245 bugs that are fixed by neither of the tools. \chatrepair's token cost on these bugs is 95.3 million and \toolname's token cost is 49.1 million. On average, \chatrepair spends 388K (95.3M/245) tokens on each of these bugs and \toolname spends 200K (49.1/245). This means \toolname saves 48\% (188K/388K) of the token cost for unfixable bugs. The token cost difference on unfixable bugs is because \toolname summarizes previously generated partial patches, while \chatrepair includes the whole previously generated patches in its new invocations.}

\modified{Moreover, \toolname spends 115K tokens on average for bugs that are only fixed by \chatrepair, while \chatrepair spends 530K tokens on average for bugs that are only fixed by \toolname.} This shows the cost difference is smaller when the bug is not fixed, compared to the bugs that are fixed. The reason is that both tools make the maximum number of LLM invocations that they are allowed to as long as they have not generated a plausible patch. 

\begin{mdframed}\noindent
    \textbf{Answer to RQ2: \rqtwo} \\
    \modified{\toolname improves the token cost by 73\% (149.4M/204.3M), compared to the baseline \chatrepair.} This is because \toolname has been designed and engineered to explore the LLM repair space with as few LLM invocations as possible.
    Also, we show that \toolname is particularly cost-efficient when it is successful at generating a correct patch for a bug, with a token total price of only 4\% on average compared to the baseline, representing a saving of 96\%. This means that future companies which will use a program repair product such as \toolname would see their program repair operational costs dramatically reduced. 
\end{mdframed}

\subsection{Results for RQ3 (Exploration)}
\label{sec:result_rq3}

\begin{table}[t]
\centering
\footnotesize
\caption{Selected bugs for studying  patch exploration.}
\label{tab:rq3_bugs}
\begin{tabular}{l c c c c}
\toprule
	BID & \#LLM\_Invocations & LEN & FPPS & PPM \\
	\midrule
    Chart\_5 & 1 & 1,452 & \xmark & \cmark \\
    Closure\_58 & 91 & 2,568 & \cmark & \xmark \\
    Closure\_128 & 1 & 263 & \xmark & \cmark \\
    Lang\_19 & 119 & 1,652 & \cmark & \xmark \\
	Lang\_53 & 101 & 5,218 & \cmark & \cmark \\
	Math\_13 & 103 & 165 & \cmark & \cmark \\
    Math\_23 & 1 & 5,380 & \xmark & \cmark \\
    Math\_89 & 1 & 95 & \xmark & \cmark \\
    Mockito\_33 & 1 & 410 & \xmark & \cmark \\
    Time\_20 & 1 & 402 & \xmark & \cmark \\
    Time\_22 & 85 & 108 & \cmark & \xmark \\
    \bottomrule
\end{tabular}
\end{table}

\begin{figure}
  \centering
  \includegraphics[width=\linewidth]{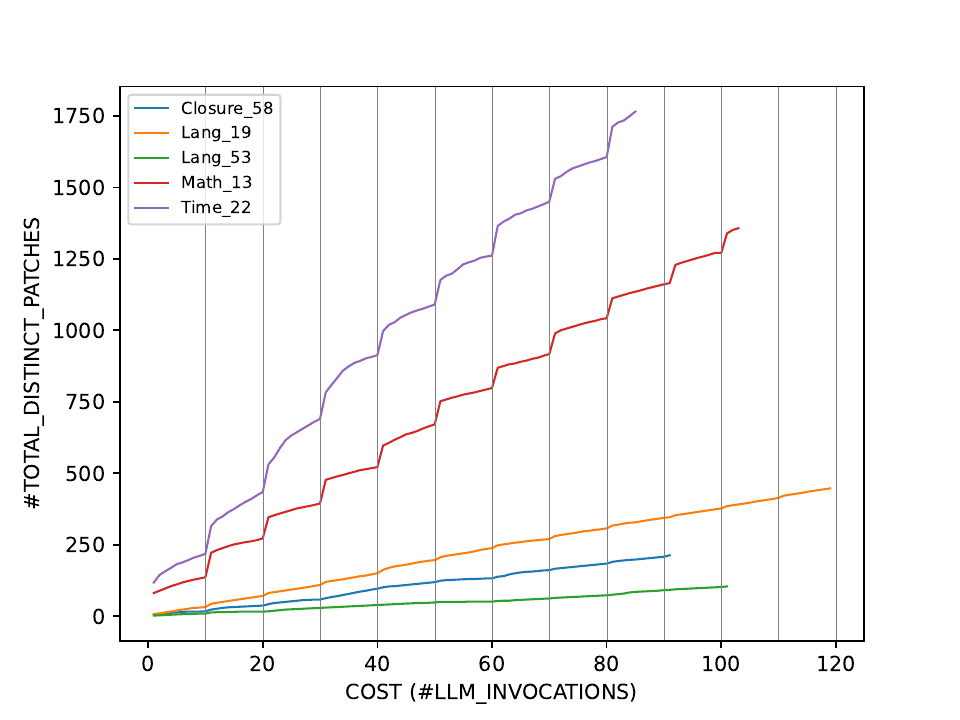}
  \caption{The progress during the plausible patch search step. The number of distinct patches increases with LLM invocations, with significant jumps at each reboot, validating the concept.}
  \label{fig:rq3_first_plausible}
\end{figure}

\autoref{tab:rq3_bugs} shows the bugs that are selected for investigating the patch exploration progress. The table shows the bug identifier (``BID''), the number of LLM invocations in the plausible search phase, and the length of the buggy function in characters (``LEN''). The ``FPPS'' and ``PPM'' columns show if the bug is considered for studying the first plausible patch phase and/or the plausible patch multiplication steps, respectively. For the first plausible patch search, we want to show how patch exploration discovers new distinct patches with reboots. Therefore, the selected bugs for studying this step are the ones with most number of reboots during the repair process. Note that no bug from ``Mockito'' and ``Chart'' is selected for this step, because all the bugs in these projects are either not fixed at all or fixed in a single round without reboots. To diversify the type of considered bugs, we also select bugs for which the first plausible patch is found with one LLM invocation. This leads to including ``Chart\_5'', ``Closure\_128'', ``Math\_23'', ``Math\_89'', ``Mockito\_33'', and ``Time\_20'' in our study of the plausible patch multiplication step.

\autoref{fig:rq3_first_plausible} shows the number of total distinct patches generated up to a certain number of LLM invocations in the first plausible patch search phase. We observe two major points in these plots. First, the number of generated distinct patches jumps after every reboot, i.e. for new rounds which are made every ten requests. After the jump at the beginning of a round, the number of new distinct patches has a slow increase until the end of that round and then jumps back when the next round starts. This indicates that the reboots are more effective than continuing a single conversation for too long. The patch exploration power of reboots with new random seed is higher.

The second observation is that the overall progress of patch exploration seems to be linear. This means at each round \toolname identifies new distinct patches and this number does not decline rapidly. Again, this corroborates the effectiveness of reboots, as they effectively lead the model to explore a different part of patch search space. For the sake of experimental costs, we were not able to discover when a potential plateau effect starts. 

\begin{figure}
  \centering
  \includegraphics[width=\linewidth]{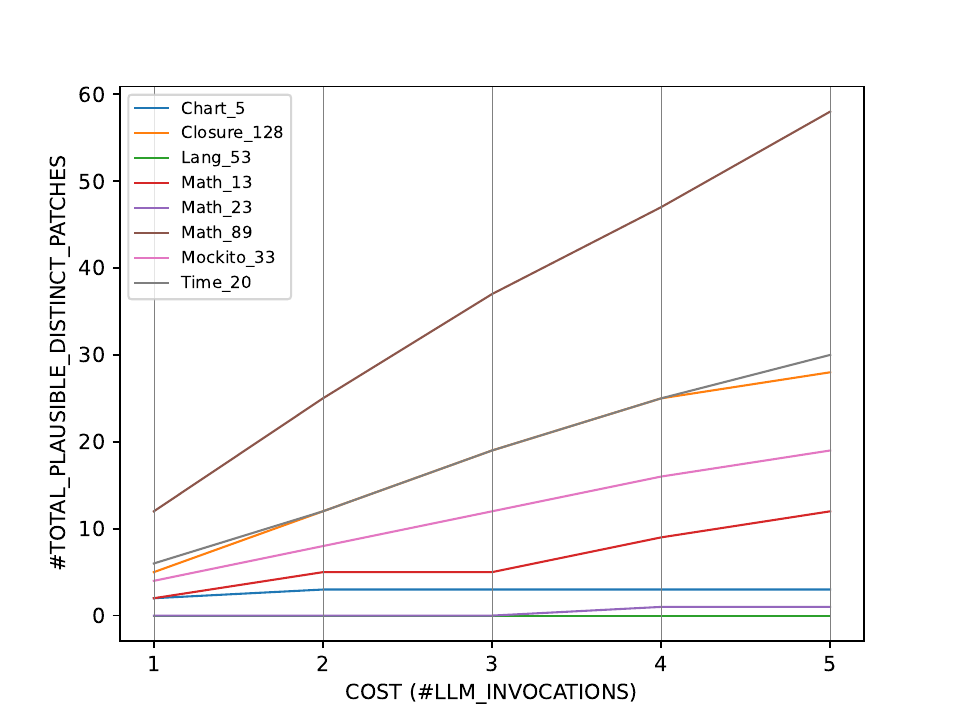}
  \caption{Progress during the plausible patch multiplication in \# distinct plausible patches. This validates the concept of plausible patch multiplication.}
  \label{fig:rq3_amplification}
\end{figure}

\autoref{fig:rq3_amplification} presents the number of total distinct plausible patches at different LLM invocation steps during plausible patch multiplication. We notice that the number of new distinct patches gradually increases. Per our experiment, five is an appropriate number of LLM invocations for patch multiplication, with this setting, \toolname takes the most advantage of patch multiplication without having invocations that yield no new distinct patches while incurring a token cost. We also see that the decline in the number of new distinct patches is generally the same over bugs. For the two longest bugs, namely ``Math\_23'' and ``Lang\_53'', the number of new distinct plausible patches is very low during patch multiplication. The reason is that \toolname generates very few patches for these two at each LLM invocation because of the token limit per LLM invocation: one cannot fit many large samples in this fixed window. The low number of distinct plausible patches for such long bugs also confirms our claim: having a limited number of invocations for plausible patch multiplication is a reasonable choice regarding the trade-off of token cost / new patches.

We also note that in both figures, the number of distinct patches is higher for bugs that are smaller, such as ``Math\_13'' and ``Time\_22'' with 165 and 85 characters, respectively. The reason is that because of the token limit per LLM invocation, \toolname can ask for more patches in each invocation given a fixed token budget.

\begin{mdframed}\noindent
    \textbf{Answer to RQ3: \rqthree} \\
    The reboot strategy in the first plausible patch search step of \toolname, enables it to efficiently explore different parts of the search space, avoiding wasting tokens in dead ends. This finding holds for various bug types: simple, complex, short, and long. In the plausible patch multiplication phase, our experiments show a real exploration of distinct plausible patches in the search space via prompting technique of patch multiplication. Both rebooting and patch multiplication are important, and put in conjunction, enable \toolname to efficiently explore the patch search space. This is significant because those original, largely unknown, prompting techniques could be used beyond program repair, in many of the software engineering tasks based on LLMs.
\end{mdframed}

\section{Threats to Validity}
\label{sec:threats}

\subsection{Construct Validity}
We consider the number of tokens used in the prompts and LLM responses to be the cost of \toolname. This is the main threat to the construct validity of this paper, since the cost can also be defined in many different ways. For example, the cost can be attributed to the number of LLM invocations, GPU usage \footnote{\url{https://huggingface.co/pricing\#spaces}}, or fine-tuning cost \footnote{\url{https://platform.openai.com/docs/guides/fine-tuning/preparing-your-dataset}}. We believe the token count is an appropriate metric for measuring the cost for two reasons. First, as explained in \autoref{sec:background}, analyzing and producing tokens are the building blocks of how LLMs work. Therefore, it is the most natural and meaningful way to measure the cost in terms of spent tokens. Second, OpenAI's models, which are the most widely used API-based LLMs, charge their users per token usage. Based on these two reasons, researchers and practitioners can safely rely on the results of our analysis of LLM-based APR cost effectiveness.

\modified{Another threat to the construct validity of our work is data leakage. The \texttt{gpt-3.5-0301} model used by \toolname is likely to have parts of the \defectsfj dataset in its training data \cite{huang2023survey}. This can lead to wrong measurement of patch exploration capabilities by LLMs. To address this threat, we follow the method used in recent related work \cite{jiang2023impact} and use another benchmark \humanevaljava. As this dataset is more recent than the training data used in \texttt{gpt-3.5-0301}, it is not prone to the data leakage problem. Our results in \autoref{sec:results_rq1} and \autoref{sec:result_rq2} show that \toolname outperforms the state-of-the-art in terms of bug-fixing performance and cost efficiency on \humanevaljava, providing evidence of external validity.}


\subsection{Internal Validity}
In our RQ1 experiment, for \textit{nl2fix} and \textsc{STEAM}, the set of considered bugs is not exactly the same as bugs considered because of the absence of publicly available reproduction packages. This poses a threat to the internal validity of our assessment of the effectiveness \toolname against state-of-the-art. However, we have reimplemented and rerun the \chatrepair on the exact same set of bugs considered in our RQ1 experiment. \chatrepair is a one of the most recent LLM-based APR tools, and it is the most relevant tool to our approach. Hence, \toolname's significant improvement over \chatrepair strongly verifies \toolname's effectiveness against the state-of-the-art.

\subsection{External Validity}
Our experiments are limited to Java projects in the \defectsfj dataset. This is a threat to the external validity of our study, as our results may not generalize to other programming languages or projects. However, \defectsfj is the most widely used and respected dataset in the area of automated program repair. Also, we have considered six different projects and various types of bugs with different lengths and complexities for our experiments. This strengthens the validity of our findings, though a more comprehensive study of the token cost of LLM-based APR in the future will certainly be valuable.

\section{Related Work}
\label{sec:related_work}

We review the previous work in three areas of research: automatic program repair with LLMs, LLM cost management, and effective repair space exploration.

\subsection{Automatic Program Repair with LLMs}
Automatic program repair is an important software engineering use case  for LLMs \cite{chen2021evaluating}. Xia and Zhang \cite{xia2022less} introduce AlphaRepair based on the CodeBert LLM \cite{feng2020codebert} and show that with an infill-style zero-shot prompt \cite{lampert2009learning} they outperform all existing learning-based APR tools. Zhang et al. \cite{zhang2023gamma} also use infill-prompts for APR, but mask the tokens based on common template-based APR patterns. Zero-shot prompts have also proven to be effective at fixing security vulnerabilities \cite{pearce2023examining}.

Recent LLM-based APR approaches, have improved the LLM's bug fixing effectiveness with prompt engineering \cite{zhang2023critical} and fine-tuning \cite{paul2023enhancing}. Cao et al. \cite{cao2023study} show that adding the code intention to the prompt and using a dialogue based approach improves the LLM's performance. Xia et al. \cite{xia2023automated} add multiple short bug-fix examples to the prompt and conduct an experiment with nine different LLMs. They conclude that the Codex LLM \cite{chen2021evaluating} outperforms other LLMs. Nashit et al. \cite{nashid2023retrieval} use an embedding-based technique, \texttt{SRoBERTa}\cite{reimers2019sentence}, and a frequency-based technique, BM-25\cite{robertson2009probabilistic}, to retrieve relevant examples for a given repair task. Similar to \cite{xia2023automated}, they show how Codex with a few-shot prompt outperforms the state-of-the-art APR tools.

Ahmed and Devanbu propose using the self-consistency approach for program repair \cite{ahmed2023better}. In this approach, the prompt contains the buggy code and the model is asked to generate multiple pairs of explanation-solutions. The prompt also contains a few examples of buggy code, explanation, and solutions. Then the solution that appears most frequently is taken as the final answer of the model. The authors show their approach outperforms the state-of-the-art on the MODIT dataset \cite{chakraborty2021multi}.

Iterative prompting strategies for APR are employed by other tools as well \cite{yuan2022circle,sobania2023analysis,fan2023automated,liu2023automated,zhang2023steam}. Most recently, Xia and Zhang introduce \chatrepair \cite{xia2023keep}, an automated iterative APR approach. \chatrepair gives the test results as feedback to the LLM to improve its patches. Once a plausible patch is generated, \chatrepair asks the LLM to generate alternative plausible patches. In contrast with \toolname, \chatrepair does not reduce the LLM costs. Namely, \toolname uses patch summarization, high sample size, and multiplication rounds to minimize the cost, while keeping strong fixing capabilities.

Researchers have also used fine-tuning to enhance LLM-based APR effectiveness \cite{haque2023fixeval,wu2023effective,zhang2023pre}. In one early work, Mashhadi and Hemmati \cite{mashhadi2021applying} fine-tune CodeBERT to automatically generated fixes for ManySStuBs4J. Zirak and Hemmati show that fine-tuning a LLMs on a dataset related to the application domain is significantly more effective than fine-tuning on an irrelevant dataset \cite{zirak2022improving}.

Jiang et al. \cite{jiang2023impact} conduct a large-scale study on using LLMs for APR. They show a fine-tuned version of InCoder \cite{fried2022incoder} that fixes 164\% more bugs compared to the state-of-the-art learning-based techniques. In another study, Huang et al. \cite{huang2023empirical} observe that fine-tuning is effective for fixing different types of bugs (test failure, vulnerability, and errors) in various programming languages (Java, C/C++, and JavaScript).

In contrast with existing LLM-based APR approaches, \toolname is the first to focus and demonstrate cost minimization without losing effectiveness. For this purpose, it uses three novel prompt strategies that have been used in previous work in the way \toolname works:
1) plausible patch multiplication;
2) summarizing feedback on previous patches; and
3) using large sample size with a high temperature.

\subsection{LLM Cost Management}
Wang et al. propose EcoOptiGen \cite{wang2023cost} to optimize the utility-cost ratio of LLMs. EcoOptiGen searches for optimal hyperparameters to generate the best solutions given a limited inference dollar budget. The considered hyperparameters include number of responses, temperature, and max tokens. The authors later introduce EcoAssistant \cite{zhang2023ecoassistant}, which uses an iterative approach to gradually improve the LLMs response. EcoAssistant first attempts to solve the problem using a cheap and weak LLM, and moves to more expensive and strong LLMs only if it is required. The authors show that EcoAssistant surpasses GPT-4's effectiveness, while reducing the cost by 50\%.

Arefeen et al. introduce LeanContext \cite{arefeen2023leancontext}. This tool reduces the LLM cost by summarizing the prompt. For this, it only includes the \emph{k} most relevant sentences from the context in the prompt. The number \emph{k} is determined using a reinforcement learning technique. In the same line of research, Mu et al. \cite{mu2023learning} propose the \emph{gisting} strategy to compress the prompt. This approach compresses the prompt into smaller `gist tokens'. Lin et al. also propose the \textsc{BatchPrompt} strategy \cite{lin2023batchprompt}, which reduces the token cost by putting more data points in a single prompt.

Chen et al. propose FrugalGPT \cite{chen2023frugalgpt}, which uses the LLM cascade strategy for cost minimization. This strategy finds the most cost-effective LLM API that can be used to solve a given problem. The authors also review two other cost minimization strategies: prompt adaptation and LLM approximation. Prompt adaptation reduces the cost by using shorter prompts. LLM approximation creates simpler and cheaper LLMs to match more expensive LLMs on a specific task.

Contrarily to other tools focused on LLM cost management, \toolname is the first system frugal LLM technique for automated program repair. \toolname contains token minimization strategies designed for and particularly suitable to APR. For example, summarizing the test results for previous patches is a unique, domain specific technique invented by \toolname.

\section{Conclusion}
\label{sec:conclusion}

In this paper, we have introduced \toolname, a novel LLM-based APR tool that focuses on minimizing the token cost. \toolname reduces the token cost by minimizing the repetitive data sent or received to LLMs and rapidly exploring various parts of the search space. More specifically, \toolname employs three strategies. 1) It summarizes previous LLM responses in new prompts and asks the LLM to generate different responses, 2) it uses reboots and patch multiplication with a high temperature setting to make sure the model looks into different parts of the search space, 3) it asks for the maximum number of responses per LLM invocation so that it explores a large part of the search space with every single prompt. \modified{Our experiments on \totalbugscnt bugs from \defectsfj and \humanevaljava shows that \toolname outperforms the state-of-the-art by fixing 33 more bugs, while reducing the token cost by 73\% (149.4M/204.3M).} For future research, our paper proposes original strategies to reduce the cost of LLMs, which can be expanded to other software engineering tasks, such as code generation.

\bibliographystyle{IEEEtran}
\bibliography{references}

\end{document}